\documentclass[onecolumn,trackchanges]{aastex631}

\usepackage{multirow}
\usepackage{booktabs}
\usepackage{amsmath,amssymb}
\usepackage[T1]{fontenc}
\usepackage{threeparttable}
\usepackage{amsmath}
\usepackage{xcolor}
\usepackage{graphicx}
\usepackage{float}
\usepackage{subfigure}
\usepackage{hyperref}
\usepackage{comment}
\usepackage{url}
\usepackage{bm}
\usepackage{epstopdf}
\defcitealias{2018ApJ...855...31W}{W18}
\shorttitle{Shear Particle Acceleration in Structured Jet for the Prompt Emission of GRB 170817A}
\shortauthors{Wang, Huang,\,\& Liang.}

\begin{document}

\title{Shear Particle Acceleration in Structured Gamma-Ray Burst Jets: II. Viewing Angle Effect on the Prompt Emission and Application to GRB 170817A}\

\correspondingauthor{Xiao-Li Huang, En-Wei Liang}
\email{xiaoli.huang@gznu.edu.cn, lew@gxu.edu.cn}

\author[0009-0001-8025-3205]{Zi-Qi Wang}
\affiliation{Guangxi Key Laboratory for Relativistic Astrophysics, School of Physical Science and Technology, Guangxi University, Nanning 530004, People’s Republic of China}

\author[0000-0002-9725-7114]{Xiao-Li Huang}
\affiliation{School of Physics and Electronic Science, Guizhou Normal University, Guiyang 550025, People’s Republic of China}
\affiliation{Guangxi Key Laboratory for Relativistic Astrophysics, School of Physical Science and Technology, Guangxi University, Nanning 530004, People’s Republic of China}

\author[0000-0002-7044-733X]{En-Wei Liang}
\affiliation{Guangxi Key Laboratory for Relativistic Astrophysics, School of Physical Science and Technology, Guangxi University, Nanning 530004, People’s Republic of China}

\begin{abstract}
Multi-messenger observations suggest that the gamma-ray burst on Aug. 17, 2017 (GRB 170817A) resulted from off-axial observations of its structured jet, which consists of a narrow ultra-relativistic jet core surrounded by a wide mild-relativistic cocoon. In a serious paper, we explore the emission of shear-accelerated electrons in the mixed jet-cocoon (MJC) region in a series of papers. This paper focuses on the viewing angle effect for a structured jet by considering the emission from the shear-accelerated electrons. It is found that the observed synchrotron emission peaks at the Infrared band and the synchrotron self-Compton (SSC) emission peaks at the band of hundreds of keV. They are not sensitive to the viewing angle. In the off-axis observations scenario, the prompt emission spectrum is dominated by the emission of the shear-accelerated electrons. The prompt gamma-ray spectrum of GRB 170817A can be well explained with our model by setting the velocity of the inner edge of the cocoon region as 0.9c, the magnetic field strength as 21 G, the injected initial electron Lorentz factor as $10^3$, and the viewing angle as 0.44 rad. We argue that the joint observations in the Infrared/optical and X-ray bands are critical to verify our model. 
\end{abstract}


\section{Introduction}
\label{sec:intro} 
Gamma-ray bursts (GRBs) are transient phenomena characterized by rapidly fluctuating high-energy electromagnetic emissions from cosmological distances \citep{1986ApJ...308L..43P,2004RvMP...76.1143P}. It has been suggested GRBs originate from ultra-relativistic jets powered by the collapse of massive stars or the merger of compact objects \citep{1992ApJ...395L..83N,1993ApJ...405..273W,2002ARA&A..40..137M,2004RvMP...76.1143P,2015PhR...561....1K}. The predicted prompt emission remains a subject of ongoing debate, encompassing a thermal component from photospheric emission and a non-thermal component arising from synchrotron (Syn) radiation and/or inverse Compton (IC) processes of accelerated electrons \citep{1986ApJ...308L..47G,1990ApJ...365L..55S,1992MNRAS.258P..41R,1993ApJ...405..278M,2001A&A...372.1071D}. One well-established phenomenological model is the on-axis "top-hat" jet scenario, which assumes a conical jet with uniform properties, and the prompt $\gamma$-ray emission is observed along the jet propagation axis \citep{2002NewA....7..197R,2004NewAR..48..459L,2019MNRAS.484L..98K}. This model has reliably replicated observational data and provided effective explanations for numerous GRBs, including GRBs 060614, 090510, and 130427A \citep{2007A&A...470..105M,2016ApJ...831..178R,2014Sci...343...48M}.

On the other hand, the role of the structured jet model has garnered attention for interpreting observed and predicted emission, particularly in the context of off-axis observations \citep{2003ApJ...591.1075K,2004ApJ...601..371L,2007ApJ...665..569M,2018MNRAS.473L.121K,2021MNRAS.500.3511G,2022MNRAS.509..903N,2023MNRAS.518.5145U}. A central focus of this investigation is the jet-cocoon structure, which has been the subject of comprehensive study \citep{2002MNRAS.337.1349R,2005ApJ...629..903L,2017ApJ...834...28N,2017ApJ...848L...6L,2019Natur.565..324I,2019ApJ...881...89L}. Fortunately, the nearby GRB 170817A maybe provide an opportunity to probe the angular structure of the jet. Previous studies have shown that the observed prompt $\gamma$-ray emission of GRB 170817A can be explained by a structured/off-axis relativistic jet with a viewing angle of $32_{-13}^{+10}\pm 1.7$ deg \citep{2017PhRvL.119p1101A,2017ApJ...848L..12A,2017ApJ...848L..13A,2018Natur.554..207M,2018ApJ...860L...2F,2018MNRAS.481.1597G,2019MNRAS.483..840B}. It has been suggested that the $\gamma$-rays should originate from at least a mild-relativistic outflow, indicating that the prompt emission could arise from a cocoon potentially formed by either a choked or successful jet\citep{2018MNRAS.479..588G,2018ApJ...867...18N}. \cite{2019MNRAS.489.1919T} presented the results of the annual afterglow monitoring of GRB 170817A, which revealed that the afterglow temporal evolution contradicted most models of choked jet/cocoon scenarios. Subsequent observations and light-curve analyses also corroborated the presence of collimated relativistic structured jet and the supported necessity of off-axis observation \citep{2019ApJ...870L..15L,2019Sci...363..968G,2021MNRAS.501.5746T,2022ApJ...927L..17H,2023MNRAS.523.4771G}. Nonetheless, a comprehensive explanation for the prompt emission physical origin of GRB 170817A remains elusive.

In off-axis scenarios, the strong relativistic beaming effect significantly suppresses the radiation contribution of the jet, impacting the validity of traditional models \citep{2002MNRAS.332..945R,2006ApJ...645.1305J,2018MNRAS.473L.121K}. This necessitates the exploration of other particle acceleration mechanisms and radiation regions within the GRB environment. Through conceptualizing the GRB ejecta as a jet-cocoon structure, we have proposed that the prompt $\gamma$-ray spectral features could potentially be explained by the combined contributions of shear-accelerated electrons in the mixed jet-cocoon (MJC) region and internal-shock-accelerated electrons in the jet core (please see \citealp{2024arXiv241111234W}). In this model, within the MJC region, characterized by a radial velocity distribution, the shear acceleration mechanism extracts energy from the background outflow to accelerate electrons, whose synchrotron self-Compton process contributes to the prompt emission \citep{1981SvAL....7..352B,1989ApJ...340.1112W,2004ApJ...617..155R,2018ApJ...855...31W,2024arXiv241111234W}. Given the mild-relativistic nature of the MJC region, relativistic beaming effects are not significant. Consequently, we propose an off-axis scenario in which the radiation from the ultra-relativistic jet is subdominant, with the prompt emission primarily arising from the radiation of shear-accelerated electrons within the MJC region. We further apply this paradigm to elucidate the prompt emission spectrum of GRB 170817A. 

In this paper, we introduce the shear acceleration model in Sec. \ref{sec:Shear acceleration}. In Sec.~\ref{sec:off-axis scenario}, we demonstrate the impact of the viewing angle on the observed radiation from both the MJC region and jet core and provided an explanation for the spectrum of GRB 170817A. The summary and discussion are presented in Sec.~\ref{sec:Summary}. Throughout this paper, we employ a Hubble constant of $H_0=71 \mathrm{km} \mathrm{s}^{-1}\,\,\mathrm{Mpc}^{-1}$, and the cosmological parameters of $\varOmega _M=0.27$ and $\varOmega _{\Lambda}=0.73$.

\section{Shear acceleration Model}
\label{sec:Shear acceleration}

We conceptualize the GRB jet-cocoon structure as comprising three distinct regions: an ultra-relativistic narrow jet core region with the constant velocity ($r<r_0$), a sub-relativistic mixed jet-cocoon (MJC) region characterized by a radially decreasing velocity ($r_0<r<r_2$), and an outer cocoon region exhibiting uniform velocity ($r_2<r$). Hereinafter, variables with the subscripts ``jet'' and ``cn" refer to the jet core and MJC regions, respectively. Building on previous research, we model the MJC region velocity profile $u_{\rm cn}(r)$ (along jet axis) as an exponential-decay function  \citep{2024arXiv241111234W}
\begin{equation}
u_{\rm cn}(r)=\beta _{\rm cn,0}e^{-k}, \ \  k = \frac{r \ln (\beta_{\rm cn,0}/\beta_{\rm cn,2})}{r_{2}}
\label{eq:u(r)}, 
\end{equation}
where $r$ represents the radial distance from the jet axis, $u_{\mathrm{cn}}$ denotes the outflow velocity normalized to the speed of light $c$, and $\beta_{\rm cn,0}$ and $\beta_{\rm cn,2}$ correspond to the fluid velocities at $r_0$ and $r_2$, respectively. Holistically, the structure is parameterized as follows: the radiation regions for both the jet core and MJC region are positioned at a distance of $R_{\mathrm{jet}} \sim 1 \times 10^{15}$ cm \citep{2002MNRAS.337.1349R,2011ApJ...726...90Z,2015AdAst2015E..22P}. The magnetic field strengths are $B_{\mathrm{jet}}\sim 10^6$ G for the jet core and $B_{\mathrm{cn}}\lesssim 10^3$ G for the MJC region \citep{2001A&A...369..694S,2006ApJ...652..482P}. The half-opening angles are defined as $\theta_{\mathrm{jet}} \sim 0.03$ rad for the jet core and $\theta_{\mathrm{cn}} \sim 0.30$ rad for the complete cocoon \citep{2015ApJ...815..102F,2019ApJ...881...89L,2023MNRAS.524.4841H}.

As described in \cite{2024arXiv241111234W}, we consider particles to be bounded within the MJC region ($r_0 < r < r_2$), where the outflow is treated as steady-state and incompressible. The shear boundary layer (SBL) serves as an electron injection site, where significant particle acceleration energizes electrons to an effective energy of $\gamma_{\mathrm{eff}}\sim \Gamma_{i}$, with $\Gamma_{i}$ denoting the jet Lorentz factor \citep{2017ApJ...847...90L}. Electrons are injected into the MJC region at $r = r_1\ (r_1 \gtrsim r_0)$ with the momentum $p_0 \sim \gamma_{e, \mathrm{inject}} / m_e c$ and subsequently experience further acceleration via the shear acceleration mechanism. 

The electron distribution function ($f_{0}$) for shear acceleration in the relativistic shear flow is described within the particle transport framework formulated by \cite{2018ApJ...855...31W}. In the strong scattering limit regime, the transport equation for the jet-cocoon structure is expressed as \citep{2018ApJ...855...31W,2024arXiv241111234W}
\begin{align}
    -\frac{1}{r}\frac{\partial}{\partial r} \left( \kappa r \frac{\partial f_0}{\partial r} \right) & -\frac{c^2}{15}\frac{\Gamma _{\mathrm{cn}}^{4}}{p^2}\left( \frac{du_{\mathrm{cn}}}{dr} \right) ^2\frac{\partial}{\partial p}\left( p^4\tau \frac{\partial f_0}{\partial p} \right) = Q, \\
    Q &= \frac{1}{2\pi r_1}\frac{N_0}{4{\pi p_0}^2}\delta \left( p-p_0 \right) \delta \left( r-r_1 \right).
\label{eq:transportequation}
\end{align}
where the particle diffusion coefficient, $\kappa$, is defined as $\kappa ={{{v}^2\tau}/{3}}$, $v$ is the particle speed in the comoving frame, and $\tau$ is the scattering or collision timescale. $p$ denotes the comoving particle momentum, $\Gamma_{\rm cn}$ is the Lorentz factor corresponding to $u_{\rm cn}$, and $Q$ represents the particle source. Then, the analytical solution for shear-accelerated electron distribution function $f_0$ can be derived as \citep{2018ApJ...855...31W}  
\begin{equation}
\begin{split}
    f_0=&\frac{15}{8 \pi^{2} \left( \xi _0-\xi _2 \right)\left| \frac{d\xi}{dr}|_{r_1} \right|r_1}\left( \frac{N_0}{{p_0}^{3}c^2\tau _0} \right) \exp \left[ -\frac{\left( 3+\alpha \right) T}{2} \right] \\
    &\times \sum_{n=0}^{\infty}{\frac{1}{y_n}\sin \left[ \left( n+\frac{1}{2} \right) \pi w_1 \right]}\sin \left[ \left( n+\frac{1}{2} \right) \pi w \right] \exp \left( -y_n\left| T \right| \right) ,
\label{eq:f0}
\end{split}
\end{equation}
in which  
\begin{equation}
\xi \left( r \right) =\frac{1}{2}\ln \left( \frac{1+u_{\mathrm{cn}}}{1-u_{\mathrm{cn}}} \right) , \ \  w \equiv \frac{\xi -\xi _2}{\xi _0-\xi _2}, \ \ T=\ln \left( \frac{p}{p_0} \right) 
\label{eq:6}
\end{equation}
\begin{equation}
y_n=\left[ \frac{5\pi ^2\left( 2n+1 \right) ^2}{4\left( \xi _0-\xi _2 \right) ^2}+\frac{\left( 3+\alpha \right) ^2}{4} \right] ^{{{1}/{2}}}, \ \  n=0,1,2,\dots ,
\label{eq:7}
\end{equation}
where the subscripts 0, 1, and 2 denote the quantities at $r=r_0$, $r=r_1$, and $r=r_2$, respectively. $\tau_0$ represents the initial scattering timescale. For the calculation, we adopt $\alpha=1/3$, in accordance with the Kolmogorov turbulence model \citep{1941DoSSR..30..301K,2002ApJ...564..291C}.

\section{Off-Axis Scenario}
\label{sec:off-axis scenario}
The electrons accelerated via the shear acceleration mechanism within the MJC region and through internal shocks in the jet core are cooled by both the Synchrotron ($\rm Syn$) radiation and the synchrotron self-Compton ($\rm SSC$) process \citep{1979rpa..book.....R}. In the off-axis Structured jet-cocoon model, the relativistic beaming effect significantly attenuates the $\gamma$-ray contribution from the ultra-relativistic jet core. As a result, the observed/predicted prompt emission may be predominantly attributed to the radiation of shear-accelerated electrons within the mild-relativistic MJC region. In this section, we calculate the isotropic-equivalent luminosity $L_{\mathrm{iso, cn}}$ for both the $\rm Syn_{cn}$ and $\rm SSC_{cn}$ emissions of shear-accelerated electrons across various viewing angles. We specify the maximum $L_{\mathrm{iso, cn}}$ at the viewing angle of $\theta_{\rm v}=0.0$ rad to $10^{50}\  \rm erg\ s^{-1}$ as the reference value. The parameters of the shear acceleration model are specified as follows: $\beta_{\rm cn,0} = 0.9$, $B_{\rm cn} = 100$ G, and $\gamma_{e, \mathrm{inject}} = 300$. 

To compare the contribution of internal-shock-accelerated electrons in the jet core, we also calculate the isotropic-equivalent luminosity $L{\rm iso, jet}$ for different viewing angles. We employ a standard broken power-law function to quantify the energy distribution of shock-accelerated electrons \citep{1978MNRAS.182..147B,2001MNRAS.328..393A}, denoted as
\begin{equation}
\frac{dN_{e,\mathrm{jet}}}{d\gamma _{e,\mathrm{jet}}}\propto \begin{cases}
	\gamma _{e,\mathrm{jet}}^{-2}&		\gamma _{\mathrm{m},\mathrm{jet}}\leqslant \gamma _{e,\mathrm{jet}}\leqslant \gamma _{\rm b,jet}\\
	\gamma _{e,\mathrm{jet}}^{-p_{\mathrm{jet}}-1}&		\gamma _{\rm b,jet}<\gamma _{e,\mathrm{jet}}\leqslant \gamma _{\rm M,jet}\\
\end{cases}. 
\end{equation}
where $\gamma_{e, \rm jet}$ denotes the electron Lorentz factor in the jet, $p_{\rm jet}$ is the spectral index of electrons accelerated via internal shocks, and $\gamma _{\mathrm{m}, \rm jet}$, $\gamma _{\rm b, jet}$, and $\gamma _{\rm M, jet}$ represent the minimum Lorentz factor, break Lorentz factor, and maximum Lorentz factor of the electrons, respectively. In this study, we set  $p_{\rm jet}=2.3$, $\gamma _{\mathrm{m}, \rm jet} = 1 \times 10^{3}$, $\gamma _{\rm b, jet} = 1 \times 10^{4}$, and $\gamma _{\rm M, jet} = 1 \times 10^{5}$, respectively.

\begin{figure}[htbp!]
    \centering
    \includegraphics[width=0.4\textwidth]{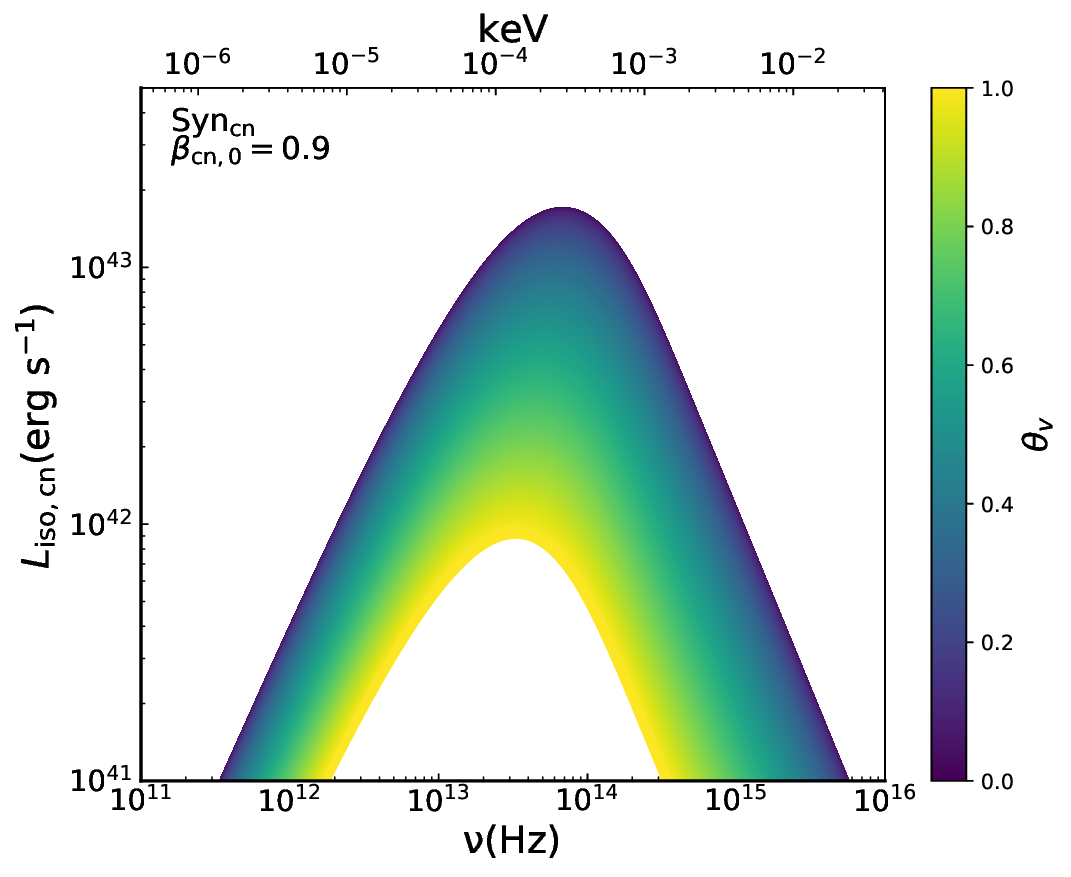}
    \includegraphics[width=0.4\textwidth]{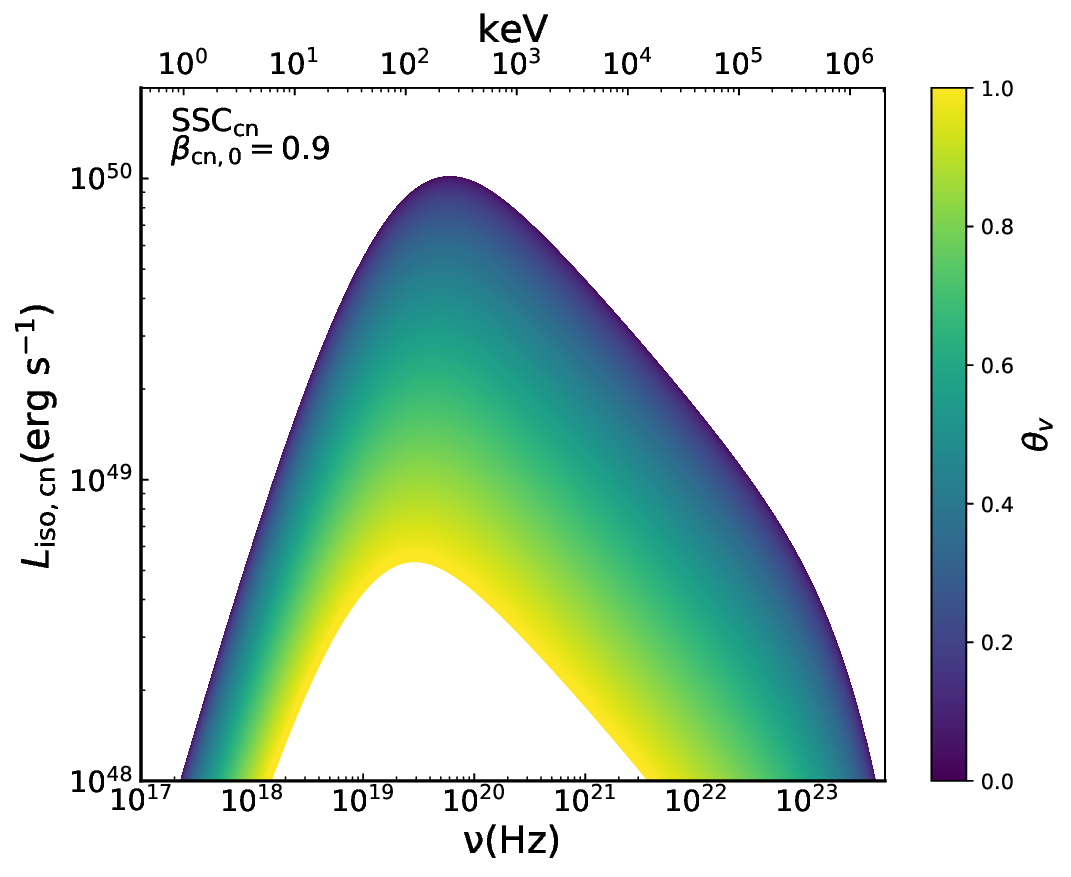}
    \caption{The isotropic-equivalent luminosity $L_{\rm iso, cn}$ of $\rm Syn_{cn}$ and $\rm SSC_{cn}$ components at various viewing angles.}
    \label{fig:cocoon09}
\end{figure}

We first present the isotropic-equivalent luminosity $L_{\mathrm{iso, cn}}$ of the $\rm Syn_{cn}$ and $\rm SSC_{cn}$ components at various viewing angles, as shown in Figure~\ref{fig:cocoon09}.
The predicted $L_{\mathrm{iso, cn}}$ of $\rm Syn$ component varies from several $ 10^{43}\,\rm erg\,s^{-1}$ to $10^{42}\,\rm erg\,s^{-1}$ across different viewing angle (left panel). The peak frequency, approximately $10^{14} \,\rm Hz$, indicates a contribution in the infrared/optical emission. Simultaneously, the inferred isotropic-equivalent luminosity $L_{\mathrm{iso, cn}}$ of the $\rm SSC_{\rm cn}$ component spans from $ 10^{50}\,\rm erg\,s^{-1}$ to $10^{49}\,\rm erg\,s^{-1}$ (right panel), with a peak frequency around  $\sim 10^{20} \,\rm Hz$. In other words, the $\rm SSC_{\rm cn}$ component produced from the shear-accelerated electrons within the MJC region dominates the luminosity. 

\begin{figure}[htbp!]
    \centering
    \includegraphics[width=0.4\textwidth]{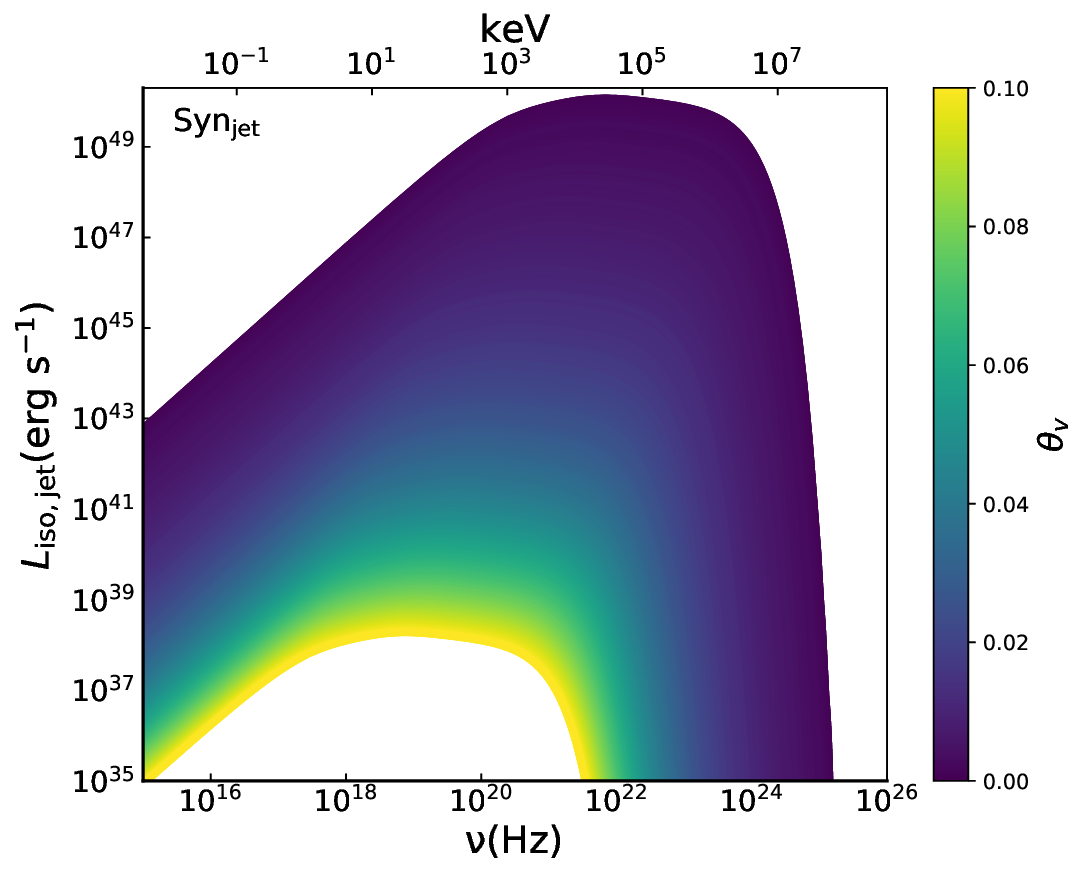}
    \includegraphics[width=0.4\textwidth]{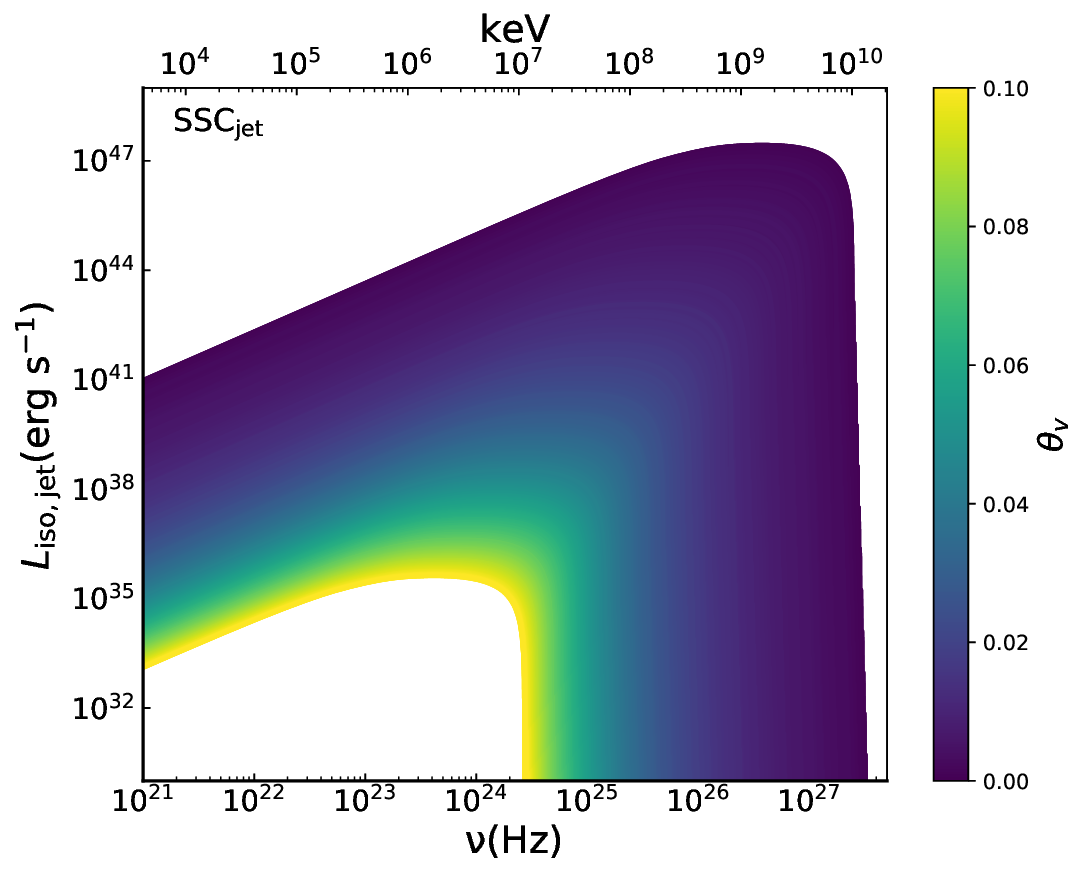}    
    \caption{The isotropic-equivalent luminosity $L_{\rm iso, jet}$ of $\rm Syn_{jet}$ and $\rm SSC_{jet}$ components at various viewing angles.}
    \label{fig:jetffaxis}
\end{figure}

Figure~\ref{fig:jetffaxis} illustrates the $L_{\mathrm{iso, jet}}$ of $\rm Syn_{jet}$ and $\rm SSC_{jet}$ components across different viewing angles.
One can find that this result is significantly different from that of the MJC region. The $\rm Syn_{jet}$ and $\rm SSC_{jet}$ emission luminosities are significantly modulated by the viewing angle owing to the ultra-relativistic beaming effect. As the viewing angle increases from $0.0$ to $0.04$ rad, the maximum isotropic-equivalent luminosity, $L_{\mathrm{iso, jet, max}}$, for both the $\rm Syn_{jet}$ and $\rm SSC_{jet}$ components experiences a dramatic decline. Specifically, $L_{\mathrm{iso, jet, max}}$ for $\rm Syn_{jet}$ ($\rm SSC_{jet}$) declines from $10^{50}\  \rm erg\ s^{-1}$ ($10^{47}\  \rm erg\ s^{-1}$) to $10^{41}\  \rm erg\ s^{-1}$ ($10^{38}\  \rm erg\ s^{-1}$), spanning up to nine orders of magnitude.

\begin{figure}[htbp!]
    \centering
    \includegraphics[width=0.32\textwidth]{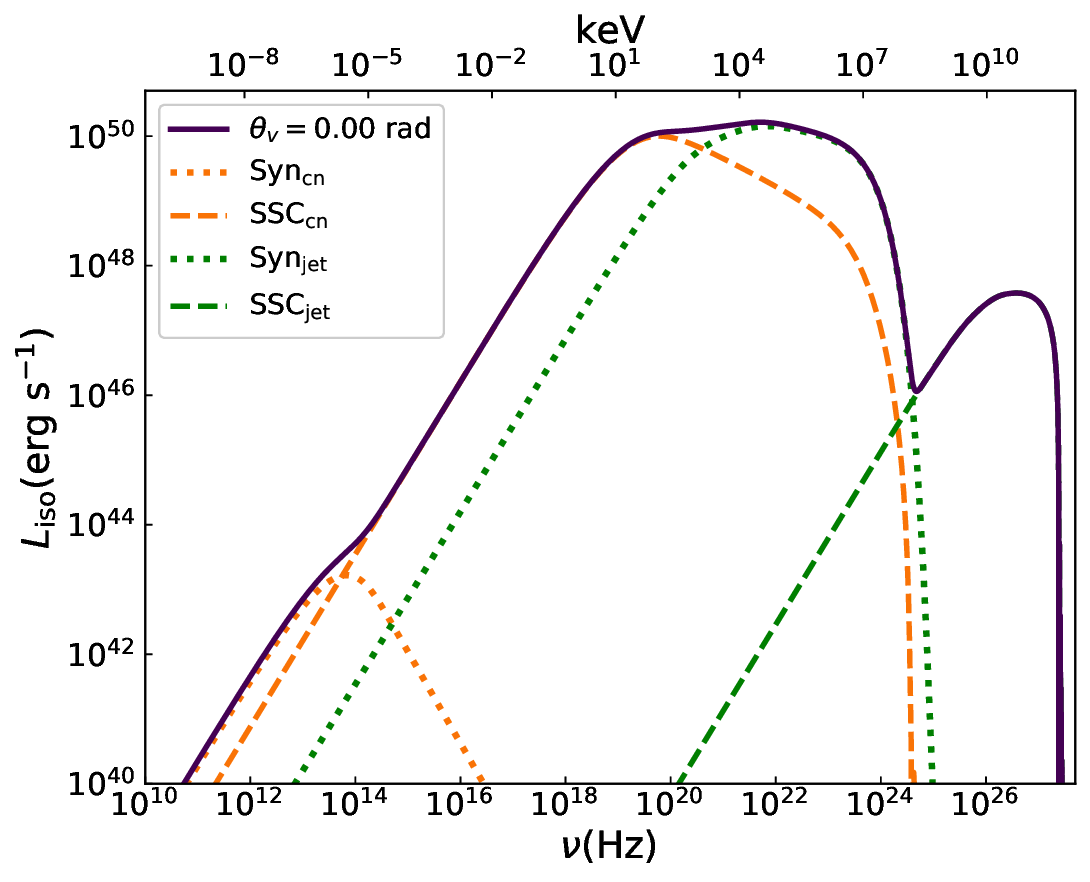}
    \includegraphics[width=0.32\textwidth]{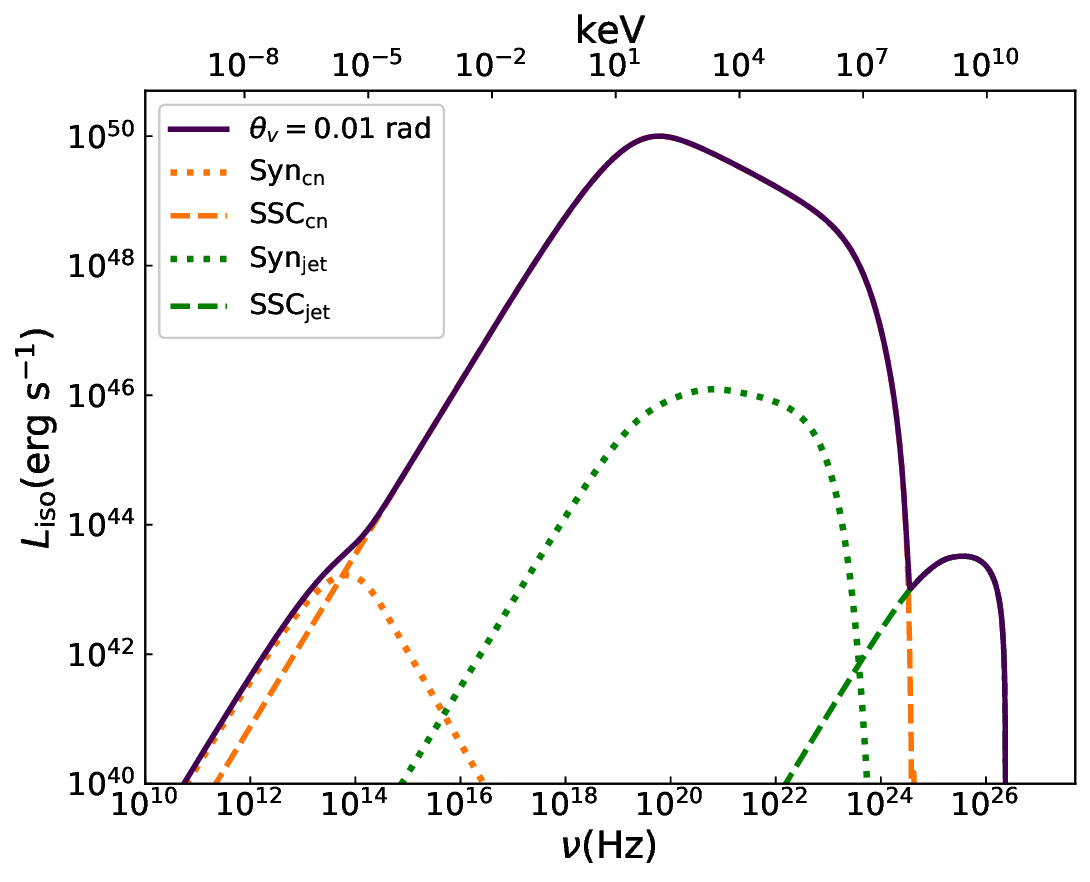}    
    \includegraphics[width=0.32\textwidth]{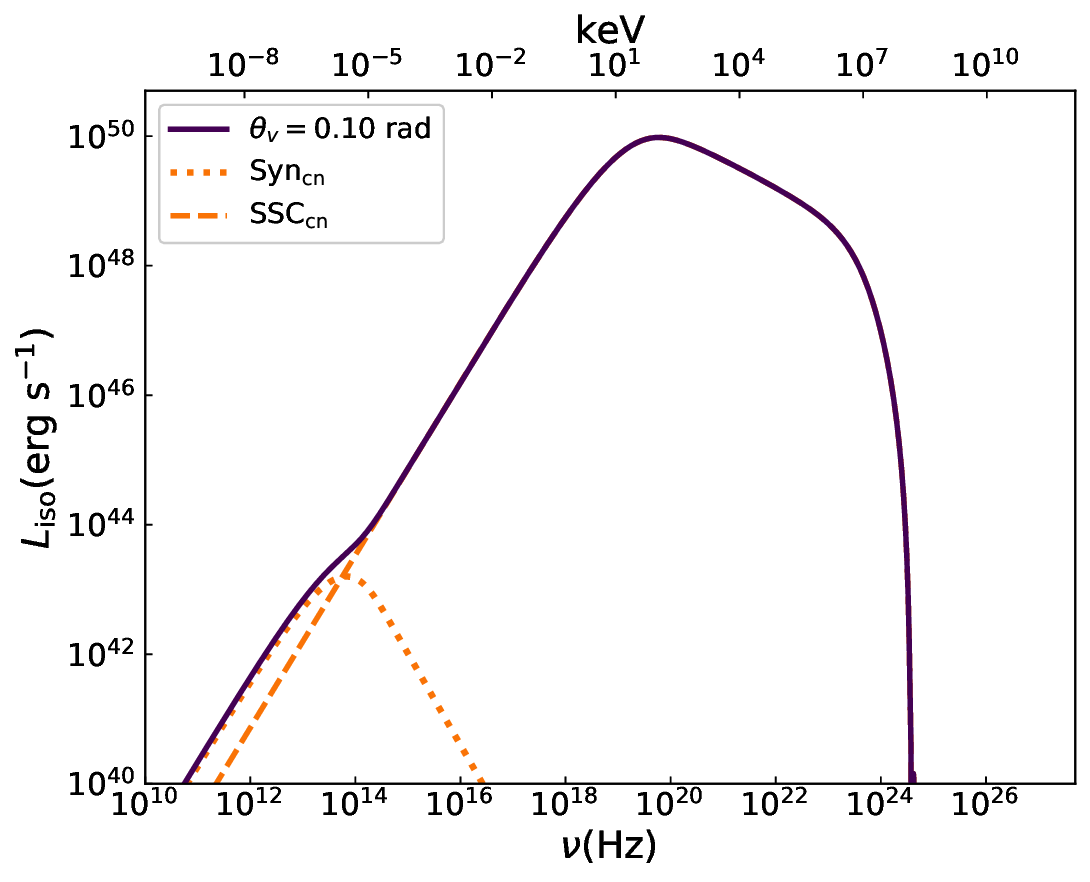}
    \caption{The schematic diagram of the isotropic-equivalent luminosity synthesized from the radiation components of the jet core and MJC region across various viewing angles $\theta_{\rm v}=0, 0.01,0.1$ rad.}
    \label{fig:synthetic spectrum}
\end{figure}

Figure~\ref{fig:synthetic spectrum} depicts the synthetic radiation spectra of the jet core and MJC region at viewing angles $\theta_{\rm v}=0,0.01,0.1$ rad. The orange and green lines represent the radiation of shear-accelerated electrons in the MJC region and internal-shock-accelerated electrons in the jet core, respectively, with the dotted and dashed lines corresponding to the $\rm Syn$ and $\rm SSC$ components in each region. Assuming a zero viewing angle to the jet core axis (left panel of Figure~\ref{fig:synthetic spectrum}), the radiation is contributed by the MJC region and the jet core. When the viewing angle is adjusted to $\theta_{\rm v}=0.01$ rad, the contribution from the jet core emission diminishes. The $\rm SSC_{\rm cn}$ component then dominates the whole predicted radiation below $\sim 10^{25}\,\rm Hz$. The result is shown in the middle panel of Figure~\ref{fig:synthetic spectrum}. In addition, the $\rm SSC_{\rm cn}$ component dominates the entire emission spectrum at a viewing angle of $\theta_{\rm v}=0.1$ rad, as shown in the right panel of Figure~\ref{fig:synthetic spectrum}. Overall, in the off-axis scenario, the effect of the viewing angle substantially suppresses the emission contribution of the jet core, allowing the radiation of shear-accelerated electrons within the MJC region to dominate the synthetic energy spectrum progressively. At sufficiently large viewing angles, the spectrum is exclusively determined by the radiation from the MJC region.

\subsection{GRB 170817A}
\label{sec:GRB 170817A}
The GRB 170817A data from the GBM reveals two components: a sharp pulse occurring between $T_0 - 0.26$ s and $T_0 + 0.57$ s, and a faint tail extending from $T_0 + 0.95$ s to $T_0 + 1.79$ s \citep{2017ApJ...848L..14G,2018NatCo...9..447ZA}. We download the GBM data (detectors NaI 1, NaI 2, and BGO 0) from the public science support center on the official $Fermi$ website\footnote{\url{http://fermi.gsfc.nasa.gov/ssc/data/}} and process the time-integrated spectrum of the main pulse period ($T_0 - 0.26$ s $-$ $T_0 + 0.57$ s) for spectral analysis within our model.
\begin{figure}[htbp!]
    \centering
    \includegraphics[width=0.5\textwidth]{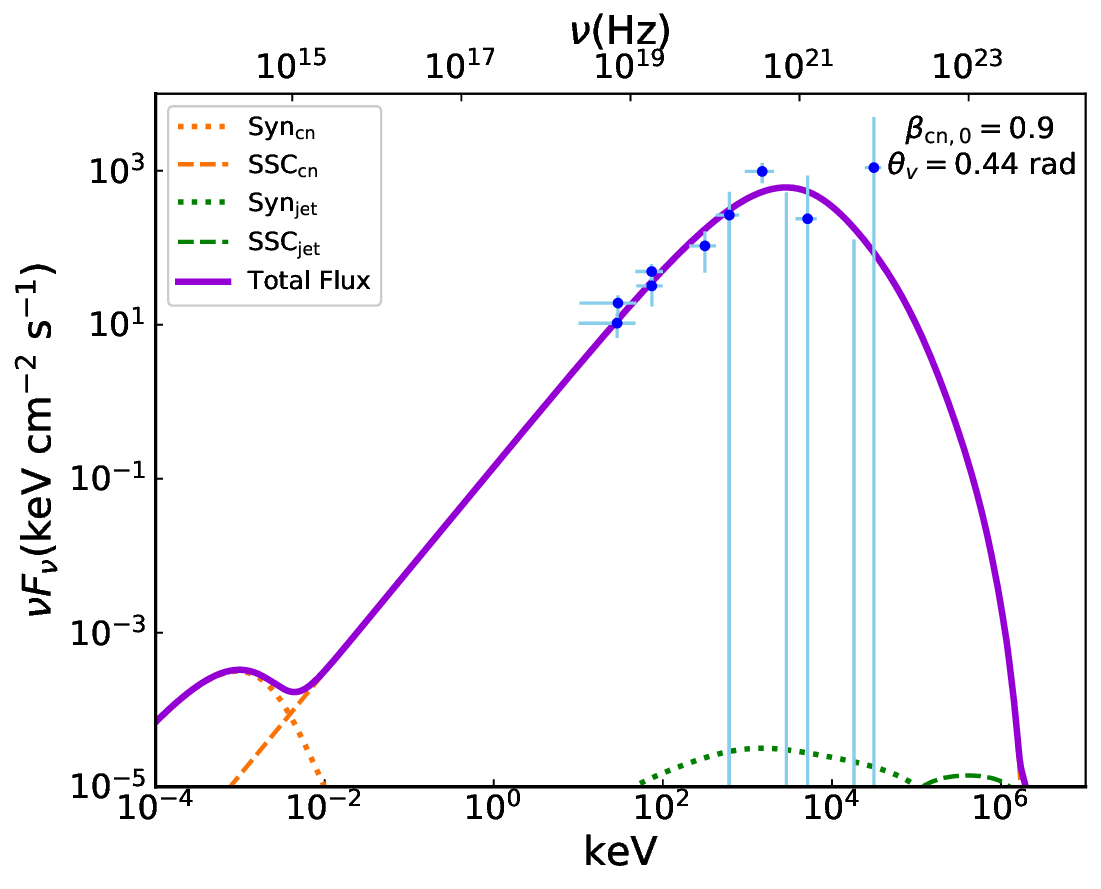}
    \caption{The time-integrated spectrum of GRB 170817A, together with the theoretical predictions from the conceptualized model with $\theta_{\rm v}=0.44$ rad (solid line). The emission components of the MJC region and the jet core are marked with dotted and dashed lines.}
    \label{fig:GRB170817A}
\end{figure}

Theoretical models of the binary neutron star mergers predict a fast ejecta tail with a velocity of $\beta_{\rm tail} \approx 0.6-0.8$, presumably originating from the interface of the merging neutron stars and plausibly associated with prompt emission \citep{2018MNRAS.475.2971B,2022ApJ...927L..17H}. Thus, we tentatively set $\beta _{\rm cn,2} \approx 0.6$. In addition, we adopt an off-axis viewing angle of $25^\circ$ (approximately $0.44$ rad) and designate the jet core as ultra-relativistic with a Lorentz factor of $\Gamma_{\rm jet} \sim 1000$, consistent with previous studies \citep{2010ApJ...716.1178A,2018Natur.554..207M,2019MNRAS.489.1919T}. The distribution parameters of shock-accelerated electrons in the jet core remain consistent with the previous text.
Figure~\ref{fig:GRB170817A} illustrates the observed data and the model fitting curve. The spectrum is effectively characterized by the formulated off-axis scenario, with the parameter set: $\beta_{\rm cn,0} = 0.9$, $B_{\rm cn} = 21$ G, and $\gamma_{e, \rm inject} = 1 \times 10^{3}$. At the large viewing angle, the $\gamma$-ray emission, spanning from keV to several hundreds of MeV, is self-consistently attributed to the $\rm SSC_{cn}$ radiation of shear-accelerated electrons within the MJC region. Conversely, the contribution of the jet core remains insignificant, even when the maximum isotropic-equivalent luminosity $L_{\mathrm{iso, jet, max}}$ is artificially elevated to transcend $10^{52} \rm \ erg\ s^{-1}$ \citep{2021MNRAS.500..627H}, as depicted by the green curve in Figure~\ref{fig:GRB170817A}. 

In our calculations, the maximum electron Lorentz factor is $\gamma_{e,\rm cn,max} \sim 3\times 10^{3}$. The cyclotron radius $r_{g}$ of energetic electrons is given by $r_{g} = pc/qB_{\mathrm{cn}}$, which yields $r_{g} \sim 5 \times 10^{12}$ cm for electrons with $\gamma_{e,\rm cn,max}$. This evaluation indicates that electrons experiencing shear acceleration can be confined within the MJC region. Furthermore, the acceleration timescale also conforms to the constraint imposed by the radiation timescale.

\section{Summary and Discussion}
\label{sec:Summary}
We propose an off-axis scenario for structured GRB ejecta, in which the prompt emission is primarily ascribed to the shear-accelerated electrons within the MJC region, with merely a marginal contribution from the jet core. We analyze the evolution of isotropic-equivalent luminosity for both $\rm Syn$ and $\rm SSC$ radiation within the MJC region and jet core, considering various viewing angles. Due to relativistic beaming effects, the contribution of shock-accelerated electrons in the jet core is significantly suppressed. In contrast, within the mild-relativistic MJC region, the radiation of shear-accelerated electrons demonstrates insufficient sensitivity to variations in the viewing angle. As a result, the radiation from MJC region of the cocoon formation can dominate the prompt $\gamma$-ray emission.

The observations of the low-luminosity GRB 170817A align with off-axis observational characteristics and suggest the presence of a mild-relativistic outflow. We correlate the predicted observation curve from our off-axis scenario with the time-integrated spectrum of the main pulse period of GRB 170817A. As presented in Figure~\ref{fig:GRB170817A}, the shear acceleration model presents an explanation for GRB 170817A, characterized by the parameters $\beta_{\rm cn,0} = 0.9$, $B_{\rm cn} = 21$ G, and $\gamma_{e, \rm inject} = 1 \times 10^{3}$. 

With the current parameter set, we initialize the jet core Lorentz factor as $\Gamma_{\rm jet } = 1000$ and the viewing angle as $\theta_{\rm v}=0.44$ rad. These preliminary values require further rigorous validation. A reduction in the Lorentz factor $\Gamma_{\rm jet }$ ($\gamma_{e, \rm inject}$) necessitates additional magnetic field energy (i.e., larger $B_{\rm cn}$) to properly regulate the peak position of the $\rm SSC_{cn}$ component. 
This analysis relies on the time-integrated spectrum, and the observed transient features probably emanate from variations in the magnetic field structure within MJC region. However, a more precise determination of the intrinsic properties of the GRB ejecta necessitates advanced numerical simulations of relativistic hydrodynamics and magnetohydrodynamics, which are beyond the scope of this paper and pose considerable challenges to achieve in the near term.

The Space Variable Objects Monitor (SVOM) mission has been successfully launched on Jun. 22, 2024. The onboard Visible Telescope (VT) operates in the visible range, specifically optimized for detecting and observing visible emissions. Its exceptional sensitivity in the red channel enables it to reach a visual magnitude of 22.5 within 300 seconds. Predictions from the shear acceleration model within the jet-cocoon structure suggest that infrared radiation with a magnitude of 17 is expected to be detected by VT, serving as a validation of the model.

\begin{acknowledgements}
This work is supported by the National Natural Science Foundation of China (Grant Nos. 12203015, 12133003). This work is also supported by the Guangxi Talent Program (“Highland of Innovation Talents”).
\end{acknowledgements}

\bibliography{sample631}{}
\bibliographystyle{aasjournal}

\end{document}